\documentstyle[aps,prb,floats]{revtex}

\begin{document}
\input epsf
\draft
\renewcommand{\floatpagefraction}{1.00}
\renewcommand{\topfraction}{1.00}
\renewcommand{\textfraction}{0.00}
\renewcommand{\bottomfraction}{1.00}
\twocolumn[\hsize\textwidth\columnwidth\hsize\csname@twocolumnfalse%
\endcsname
\title{
  {\rm\small\hfill (submitted to Phys. Rev. B})\\
Metastable precursors during the oxidation of the Ru(0001) surface}

\author{Karsten Reuter$^{1}$, M.Ver{\'o}nica Ganduglia-Pirovano$^{1}$,
Catherine Stampfl$^{1,2}$, and Matthias Scheffler$^{1}$}

\address{$^1$Fritz-Haber-Institut der Max-Planck-Gesellschaft, Faradayweg 4-6,
D-14195 Berlin, Germany}

\address{$^2$Northwestern University, 2145 Sheridan Road, Evanston IL60208, USA}

\date{Received 19 September 2001}

\maketitle

\begin{abstract}
Using density-functional theory, we predict that the oxidation of
the Ru(0001) surface proceeds via the accumulation of sub-surface
oxygen in two-dimensional islands between the first and second
substrate layer. This leads locally to a decoupling of an O-Ru-O
trilayer from the underlying metal. Continued oxidation results 
in the formation and stacking of more of these trilayers, which
unfold into the RuO${}_2$(110) rutile structure once a critical
film thickness is exceeded. Along this oxidation pathway, we
identify various metastable configurations. These are found to be
rather close in energy, indicating a likely lively dynamics between
them at elevated temperatures, which will affect the surface
chemical and mechanical properties of the material.
\end{abstract}

\hfill {\quad}
]

\section{Introduction} 

The interaction of metals with our oxygen-rich atmosphere leads to
the oxidation of the metal surfaces. Although this is common knowledge,
little is known about the {\em microscopic processes} that actuate
this oxide formation. Roughly speaking, the reaction sequence may be
divided into the initial dissociation of O${}_2$ and O chemisorption,
followed by oxide nucleation, and finally the growth of the formed
oxide film. In this scheme, particularly the transition from a
two-dimensional on-surface O adlayer to the three-dimensional 
surface-oxide nucleus has hitherto barely been addressed.

Based on a host of density-functional theory (DFT) calculations, we
present an atomistic pathway for the oxide formation on the
Ru(0001) surface \cite{reuter01a}. The situation for this surface is in
fact unique, as both the initial O chemisorption on the metal and the
finally resulting RuO${}_2$(110) oxide patches have already been 
characterized experimentally on an atomic level \cite{menzel99,over00}.
Bridging this detailed knowledge of the initial and final state of the
oxidation, we predict that after the completion of a full monolayer
of chemisorbed O on Ru(0001), the incorporation of O into the lattice 
leads to the formation of two-dimensional sub-surface O islands between
the first and second substrate layer. This implies that domains are
formed that have a local $(1 \times 1)$ periodicity and that can be
described as a trilayered O${}_{\rm ad}$-Ru-O${}_{\rm sub}$ film on top
of Ru(0001). Further O incorporation also occurs between the first and
second substrate layer, saturating the underlying metal and almost
completely decoupling the O-Ru-O trilayer. The on-going oxidation
results in the successive formation of more of these O-Ru-O trilayers, 
which at first remain in a loosely coupled stacking sequence. Once a
critical film thickness is exceeded, this trilayer stack unfolds into
the experimentally reported RuO${}_2$(110) rutile structure \cite{over00}.

This suggested oxidation pathway proceeds via several metastable states
before the final bulk RuO${}_2$ oxide structure is attained.  Their
experimental characterization would therefore provide the ultimate
confirmation of our theoretical model. However, as these intermediate 
configurations are only metastable, their experimental identification
will be challenging, requiring sophisticated choices of temperatures
and pressures. Moreover, we find these structures to be very close in
energy to the final bulk oxide. This points at a likely lively dynamics
between these metastable configurations under realistic conditions, which
might affect the chemical and mechanical properties of the surface.

\section{Theoretical}

Our DFT calculations employ the generalized gradient approximation (GGA)
of the exchange-correlation functional \cite{perdew96}, using the
full-potential linear augmented plane wave method (FP-LAPW)
\cite{blaha99,kohler96,petersen00} to solve the Kohn-Sham equations.
The Ru(0001) surface is modeled using a slab with six metal layers
for oxygen coverages up to two monolayers (ML) and a 10 metal layer
slab for higher coverages. O is adsorbed on both sides, fully relaxing the
outermost two (three) Ru layers of the six (ten) layer slabs, as well
as the position of all O atoms. A vacuum region corresponding to five
Ru interlayer spacings ($\approx$11{\AA}) is employed to decouple the
surfaces of consecutive slabs in the supercell approach. The calculated
geometries of the four ordered adlayers of O on Ru(0001) 
($0 < \theta \le 1$ ML)  are in very good agreement with existing LEED
analyses \cite{menzel99,stampfl96a}, as well as with earlier DFT
pseudo-potential calculations \cite{stampfl96b,stampfl99}.

The FP-LAPW basis set is taken as follows: $R_{\rm{MT}}^{\rm{Ru}}=$2.3 bohr,
$R_{\rm{MT}}^{\rm{O}}=$1.3 bohr, wave function expansion inside the muffin
tins up to $l_{\rm{max}}^{\rm{wf}} = 12$, potential expansion up to
$l_{\rm{max}}^{\rm{pot}} = 4$, and local orbitals for the $4s$ and $4p$
semi-core states of Ru. The Brillouin zone integration for the
$(1 \times 1)$ cells is performed using a $(12 \times 12 \times 1)$
Monkhorst-Pack grid with 19 {\bf k}-points in the irreducible part. For the
larger surface cells, the grid is reduced accordingly, in order to obtain
the same sampling of the reciprocal space. The energy cutoff for the plane
wave representation in the interstitial region between the muffin tin spheres
is 17 Ry for the wave functions and 169 Ry for the potential.

The central quantity we obtain from the calculations is the average
binding energy of oxygen defined as

\begin{equation}
E_{\rm b}(\theta) \;=\; -\frac{1}{N_{\rm O}} \;\left[\; 
E^{\rm slab}_{\rm cov.} - E^{\rm slab}_{\rm clean} - 1/2 N_{\rm O} E^{\rm mol.}_{\rm O_2}
\;\right],
\label{bindeng}
\end{equation}

\noindent
where $N_{\rm O}$ is the total number of O atoms (on- and sub-surface)
present in the unit-cell at the considered coverage, $\theta$.
$E^{\rm slab}_{\rm cov.}$, $E^{\rm slab}_{\rm clean}$, and $E^{\rm mol.}_{\rm O_2}$
are the total energies of the slab containing oxygen, of the corresponding
clean Ru(0001) slab, and of an isolated oxygen molecule respectively.
Thus a positive binding energy, $E_{\rm b}(\theta)$, indicates that
the dissociative adsorption of O${}_2$ is exothermic.

Due to the small bond length, the total energy of the isolated O${}_2$ 
molecule, $E^{\rm mol.}_{\rm O_2}$, cannot directly be calculated
with the muffin tin radius, $R_{\rm{MT}}^{\rm{O}}=$1.3 bohr, chosen
for the surface calculations. Correspondingly we compute the total
energy of an isolated oxygen atom with $R_{\rm{MT}}^{\rm{O}}=$1.3 bohr
inside a cubic cell of side length 15 bohrs with $\Gamma$-point 
sampling of the Brillouin zone without spherically averaging the 
electron density in the open shell. To arrive at $1/2 E^{\rm mol.}_{\rm O_2}$,
we then add to the atomic energy one half of the binding energy
of the O${}_2$ molecule. The latter is calculated inside the same
box using oxygens with $R_{\rm{MT}}^{\rm{O}}=$1.1 bohr, where due
to the smaller muffin tin radius the kinetic-energy cutoff for the
plane-wave basis needed for the interstitial region has been increased
to 24 Ry \cite{ganduglia99}.

The numerical accuracy of $E_{\rm b}(\theta)$ is limited by the finite
FP-LAPW basis set, as well as by the finite size of slab and vacuum
region in the supercell approach. To check this accuracy we selectively
increased these parameters and compared the binding energies within a subset
of the considered structures. This subset is formed by the two most stable
of all tested geometries at each full monolayer coverage in the discussed
$0 \le \theta \le 3$ ML sequence, i.e. the clean surface at $\theta = 0$\,ML
and the two most stable O sites at $\theta = 1, 2$ and 3\,ML.
As the main error is introduced by the DFT description of the atomic and
molecular oxygen, a possible error cancelation will be least effective for
the absolute value of the binding energy, which however is not the quantity
entering our physical argument. Rather, it is the difference in binding
energies of two geometries, which determines which one is more stable
(if they contain the same amount of oxygen) or how $E_{\rm b}(\theta)$
evolves with coverage (if the structures contain an unequal amount of
O). The accuracy will be largest in the former case, where the
molecular energy entering eq. (\ref{bindeng}) completely cancels, so that
we will always state two values indicating the numerical uncertainty, 
when comparing relative binding energies in geometries containing an equal
(unequal) amount of O.

To assess the quality of the basis set, we increased the plane
wave cutoff in the interstitial region from 17 Ry to 24 Ry, which
leads to $\pm 5$ meV ($\pm 10$ meV) changes in the binding energy
differences within the considered subset of geometries. Use of a
denser $(20 \times 20 \times 1)$ mesh with 44 {\bf k}-points in the
irreducible wedge resulted in similar variations. Further increasing
the potential angular expansion parameter to $l_{\rm{max}}^{\rm{pot}} = 6$
and the plane wave cutoff of the potential representation to 256 Ry 
hardly affected the relative binding energies at all ($\pm 3$ meV,
$\pm 5$ meV). A similar result was obtained, when increasing the vacuum
region to 19 {\AA}. The error due to the finite slab size is the larger,
the more sub-surface oxygen is present in the slab. We repeated all
calculations for the $(1 \times 1)$ geometries at 0, 1, and 2 MLs done
originally with a six metal layer slab, now using ten metal layer slabs.
While the absolute binding energies were lowered by up to 60\,meV,
the obtained relative differences in $E_{\rm b}(\theta)$ between
the different tested geometries were within $\pm 20$ meV ($\pm 30$ meV). 
Combining all these tests, we can give a conservative estimate of the
numerical uncertainty of $\pm 30$\,meV ($\pm 50$\,meV), when comparing 
relative binding energies in geometries containing an equal (unequal)
amount of O. This inaccuracy never influences the energetic sequence 
among the structures, i.e. the most stable geometry at one coverage
always remains the most stable one, neither does it change the trend
of $E_{\rm b}(\theta)$ with coverage. Hence, none of the physical 
conclusions drawn are affected by the numerical error bars.

\section{Results}

\subsection{Oxygen chemisorption and initial incorporation}

\begin{figure}
\epsfxsize=0.48\textwidth \centerline{\epsfbox{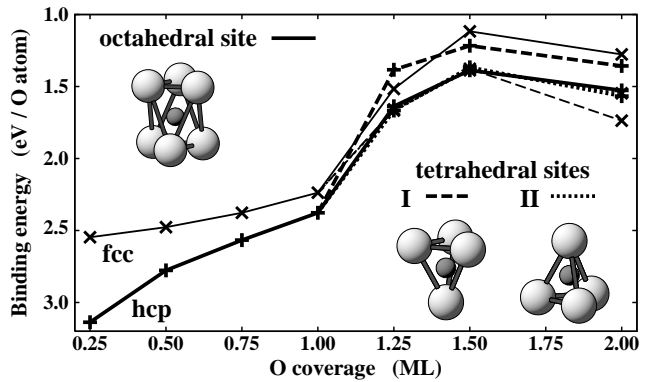}}
\caption{Binding energies per O atom, $E_{\rm b}(\theta)$, with respect
to 1/2 O${}_2$, cf. eq. (\ref{bindeng}). Coverages with $\theta < 1$\,ML
correspond to pure on-surface adsorption (hcp or fcc site). For
$\theta > 1$\,ML an O(1$\times$1) arrangement is always present at 
the surface, while the remaining O is located in either the octahedral 
or one of the two tetrahedral sites between the first and second 
substrate layer. The inserts show the local atomic coordination of 
each of these sub-surface sites (Ru = light, big spheres, O = dark, small
spheres). From the six possible structures at each coverage, the
three with the on-surface O in hcp (fcc) sites are drawn with thicker
(thinner) lines.}
\label{bindenergies}
\end{figure}

\begin{figure}
\epsfxsize=0.48\textwidth \centerline{\epsfbox{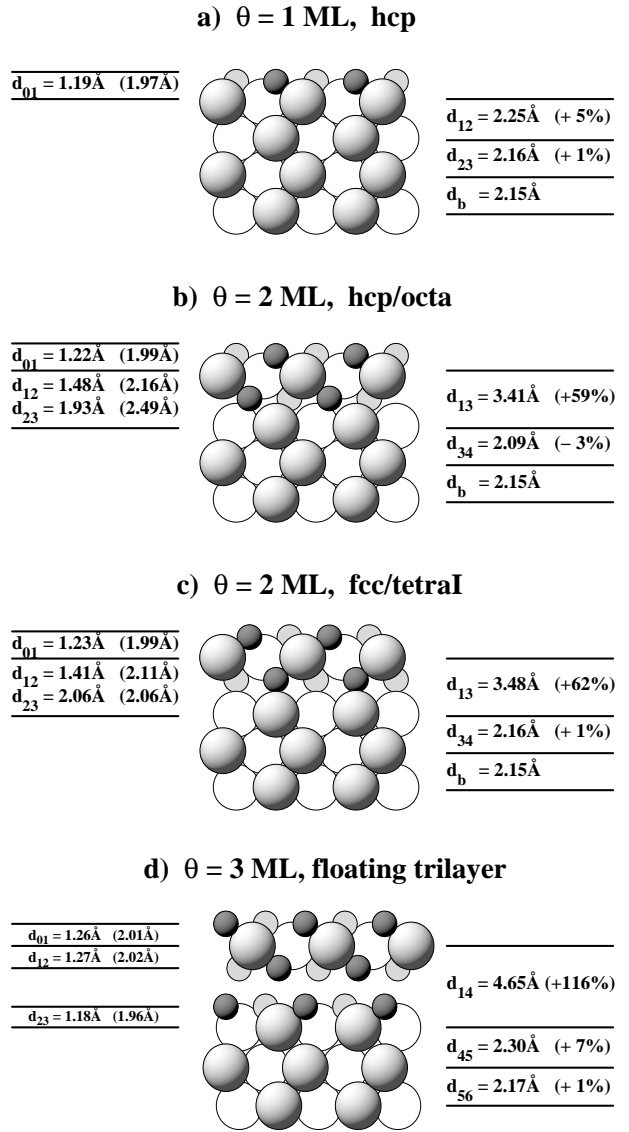}}
\caption{Sideview of several geometries along the oxidation
pathway (see text). Ru = light big spheres, O = dark, small
spheres, atoms not lying in the plane itself are white.}
\label{structures}
\end{figure}

The initial O chemisorption on the clean Ru(0001) surface proceeds via
the formation of four ordered adlayer phases, in which with increasing
coverage oxygen consecutively fills the four available hcp sites inside
a $(2 \times 2)$ unit cell \cite{menzel99,stampfl96b}. Although the
calculated binding energy, cf. Fig. \ref{bindenergies}, decreases
markedly during this coverage sequence, $0 < \theta \le 1$\,ML, the
formation of a complete monolayer on the surface corresponding to
the O$(1 \times 1)$/Ru(0001) phase shown in Fig. \ref{structures}a
is still highly exothermic \cite{stampfl96a,stampfl96b}. It is
worth mentioning that despite this decrease in the binding energy,
the calculated O-Ru bond length remains almost constant, varying only from
2.01 {\AA} in the O$(2 \times 2)$ to 1.97 {\AA} in the O$(1 \times 1)$
phase. We find the available sub-surface sites to be significantly
less favourable compared to the on-surface sites, so that we can safely
rule out O incorporation into the lattice before the full chemisorbed
overlayer is formed \cite{stampfl99,todorova01}. A similar conclusion
has recently been reached in a thorough experimental study by B\"ottcher
and Niehus \cite{boettcher99}.

To address the ensuing O penetration into the Ru(0001) surface we first
consider all available high-symmetry sub-surface sites between the first
and second substrate layer. Namely, these are one octahedral site with
sixfold Ru coordination (henceforth referred to as octa) and two tetrahedral
sites with fourfold Ru coordination (henceforth referred to as tetra-I
and tetra-II). The octa and tetra-I sites are the sites directly
below the on-surface fcc and hcp sites respectively, whereas the
tetra-II site is located below a surface Ru atom as sketched in
Fig. \ref{bindenergies}. A coverage beyond 1\,ML is obtained by combining
the O$(1 \times 1)$ adlayer in either hcp or fcc sites with the surplus
oxygen in one of the three available sub-surface sites. This results
in a total of six possible geometries at each coverage considered.
To obtain the trend of $E_{\rm b}(\theta)$ in the coverage range
$1 < \theta \le 2$\,ML we employ $(2 \times 2)$ unit cells and
calculate the binding energy of geometries with one
($\theta = 1.25$\,ML), two ($\theta = 1.50$\,ML) and four
($\theta = 2.00$\,ML) sub-surface oxygens.

The corresponding average binding energies, $E_{\rm b}(\theta)$
for 1\,ML $< \theta \le 2$\,ML, are markedly lower than the ones in the
chemisorption regime for any of the possible site combinations as
shown in Fig. \ref{bindenergies}. This reflects the aforementioned
fact, that the now additionally occupied sub-surface sites are
considerably less favourable. Nevertheless, all of the structures
are still exothermic with respect to clean Ru(0001) and molecular
oxygen, and should thus be able to form. Strikingly, $E_{\rm b}(\theta)$
for 1\,ML $< \theta \le 2$\,ML does not decrease monotonically with
coverage anymore, but shows an inflection, so that the most stable
configuration in this coverage range is found for a 2\,ML geometry
with the on-surface O in fcc and the sub-surface O in tetra-I sites,
cf. Fig. \ref{bindenergies} and \ref{structures}c. This implies
that it is energetically more favourable to place a given amount
of sub-surface oxygens in a finite area at the high local
2\,ML coverage into this fcc/tetra-I geometry, rather than to
distribute the same number of sub-surface oxygens homogeneously over
the surface, where then only the lower binding energy corresponding
to the lower local coverage is gained. Contrary to the chemisorption 
regime, where the decreasing binding energy with coverage indicates
that a repulsive interaction between the electronegative adsorbates
favours the formation of sparse ordered adlayers, we thus find a
tendency for the sub-surface oxygen to accumulate in dense two-dimensional
islands with a local $(1 \times 1)$ periodicity. To verify that these
islands are indeed two-dimensional, i.e. that the sub-surface oxygen
accumulates only between the first and second substrate layer, we
additionally calculated a series of geometries, where the coverage
between the first and second metal layer was smaller than one and
an additional oxygen atom was placed in sites between the second
and third metal layer. All of these cases were significantly less
stable than the 2\,ML fcc/tetra-I geometry, so that we can safely
exclude a dendritic 3D growth of the sub-surface O islands into
the substrate.

In these two-dimensional islands, the incorporated oxygen induces a
significant distortion of the metal lattice. The first layer distance
is expanded by 62\% compared to the Ru bulk value, so that instead of
an oxygen embedding in the host the geometry is more appropriately
described in terms of an O${}_{\rm ad}$-Ru-O${}_{\rm sub}$ trilayer
on top of a Ru(0001) substrate, cf. Fig. \ref{structures}c. It is 
worth mentioning that we find similarly large deformations for all
of the tested structures: Even inside the crystal the calculated O-Ru
bondlengths are always close to $\sim$ 2.1 {\AA} for O in tetrahedral
sites and around 2.2 {\AA} for O in octahedral sites. These optimum
bondlengths are incompatible with the available geometric space in 
the tetrahedral or octahedral interstitial sites of the Ru lattice,
which would only allow for bondlengths of 1.65 {\AA} or 1.90 {\AA}
respectively. Hence, O incorporation (particularly into the tetrahedral
sites) invariably leads to a strong local deformation of the metallic
lattice.

As such distortions are easier established at the surface, this
would then render oxygen incorporation deeper into the bulk less
favourable. Indeed, we find by more than 0.1\,eV decreased binding
energies at total coverages of $\theta = 1.25$\,ML and
$\theta = 2.00$\,ML, when placing O into the three sub-surface sites
between the second and third substrate layer compared to the hitherto
discussed incorporation directly below the surface, i.e. between
the first and second substrate layer. To address the possibility
of bulk dissolved oxygen, we also calculate the binding energy of one
O in an octahedral interstitial site in a large $(4 \times 4 \times 4)$
Ru bulk unit-cell with 64 metal atoms, where we allow its nearest Ru 
neighbours to relax. Due to the finite size of the used supercell, 
the resulting $E_{\rm b} = - 1.76$\,eV does not take the long-range
elastic interactions of the metallic lattice properly into account.
We estimate that the latter would not improve the binding energy
by more than 0.5\,eV, so that we may state a conservative upper
limit of $E_{\rm b} \approx -1.25$\,eV for the binding energy of
bulk dissolved oxygen. Although the latter species is therefore
energetically significantly less stable than the afore discussed
sub-surface O, just the vast number of available sites might still 
allow to deposit considerable amounts of oxygen into the sample
at finite temperatures. To test this, the concentration, 
$N_{\rm O}/N_{\rm sites}$, of $N_{\rm O}$ oxygens in the $N_{\rm sites}$
available sites in the crystal can be estimated by minimizing the
Gibbs free energy, which leads to \cite{ashcroft76}

\begin{equation}
\frac{N_{\rm O}}{N_{\rm sites}} \;=\; e^{E_{\rm b}/k_B T},
\end{equation}

\noindent
where $k_B$ is the Boltzmann constant. Inserting our upper limit
for the binding energy, the concentration would still be as low as
$10^{-21}$ and $10^{-8}$ at room temperature and $T = 800$\,K
respectively. As in surface science experiments uptakes are usually
quantified in ML, we convert these concentrations by considering
a cubic Ru crystal of 1 cm${}^3$ volume. The number of octahedral
interstitial sites inside Ru is $7.25 \cdot 10^{22}$ sites/cm${}^3$
and a coverage of 1\,ML corresponds to $1.56 \cdot 10^{15}$
atoms/cm${}^2$. Hence, the above stated concentrations would translate
into the equivalents of $10^{-14}$\,ML and 0.5\,ML at room temperature
and $T = 800$\,K respectively. From this we conclude, that the total
amount of bulk dissolved oxygen in a Ru crystal is completely negligible
in the considered temperature range and that incorporated oxygen
will predominantly stay as close as possible to the surface.

In conclusion, these results sketch the following picture of the
initial oxidation of the Ru(0001) surface: After the completion of
a dense O$(1 \times 1)$ adlayer on the surface, subsequent O
incorporation occurs preferentially into sites directly below the
surface. This sub-surface O accumulates in two-dimensional islands
with a local $(1 \times 1)$ periodicity, leading to the formation of
an O-Ru-O trilayer on top of the Ru(0001) substrate. In this
trilayer, the oxygens occupy fcc and tetra-I sites on and below the
surface respectively, in contrast to the coexisting O$(1 \times 1)$
domains, where the on-surface oxygen sits in hcp sites.

\subsection{Trilayer formation and surface registry shift}

The already mentioned perception of the geometry in the sub-surface
islands as an O${}_{\rm ad}$-Ru-O${}_{\rm sub}$ trilayer on top of a
Ru(0001) substrate is also what yields to an understanding of the particular
stability of the 2\,ML fcc/tetra-I structure. Focusing first on the
internal trilayer geometry, we notice that only the combinations hcp/octa
and fcc/tetra-I lead to O octahedra formed by three on-surface and
three sub-surface oxygens surrounding each Ru atom, cf. Fig. \ref{structures}b
and c. This sixfold oxygen coordination of Ru, similar to the situation in
the RuO${}_2$ bulk oxide, is energetically preferred to the fourfold
coordination with the Ru atoms located in a tetrahedral configuration,
present in the two geometries with the sub-surface O in tetra-II positions.
Although the two remaining possibilities, fcc/octa and hcp/tetra-I, would
also result in a sixfold O coordination of the metal atoms, they are
electrostatically considerably less favourable, as there the electronegative 
on- and sub-surface oxygen atoms sit directly on top of each other at
rather close distance.

\begin{figure}
\epsfxsize=0.48\textwidth \centerline{\epsfbox{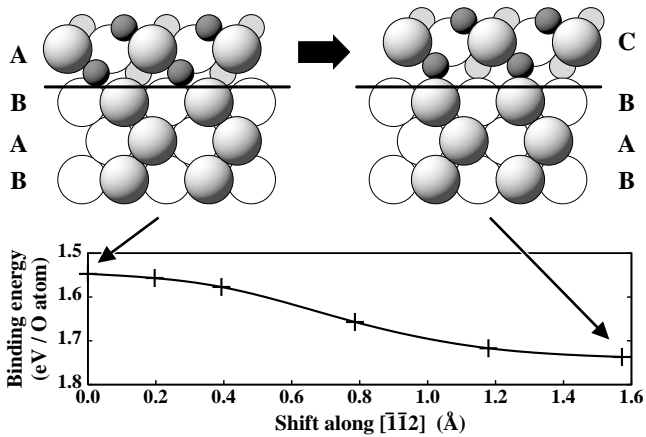}}
\caption{Atomic geometries and binding energy during the
trilayer shift along the $[\bar{1}\bar{1}2]$ direction.
Starting from the hcp/octahedral configuration (top left),
the trilayer slides over the Ru(0001) substrate and ends
up in a fcc/tetrahedral-I arrangement (top right), where
the first Ru layer lies in a stacking fault position
(Ru = light, big spheres, O = dark, small spheres,
atoms not lying inside the plane itself are white).}
\label{sliding}
\end{figure}

While the trilayer geometry itself thus favours the hcp/octa and
fcc/tetra-I configurations, it is the coupling to the underlying substrate
that finally leads to the preference for the latter combination. Actually,
the prior geometry is in fact unstable against a registry shift of the
whole trilayer along the $[\bar{1}\bar{1}2]$ direction as drawn in Fig.
\ref{sliding}. At the end of this barrierless displacement, in which the
complete O${}_{\rm ad}$-Ru-O${}_{\rm sub}$ film slides over the Ru(0001)
surface, the on-surface oxygens are located in fcc and the sub-surface O in
tetra-I sites. As both the electronic and the geometric structure
within the trilayer are found to remain virtually unchanged during this
motion, we attribute the substantial binding energy gain of 0.2\,eV per
O atom primarily to an improved trilayer-substrate bonding between
the sub-surface oxygens and the topmost Ru atoms of the underlying
metal (Ru${}_{\rm 2nd}$). This is also corroborated by the outward movement
of the latter, such that the corresponding layer distance, $d_{34}$, relaxes
from a clean-surface like $-3$\% contraction in the hcp/octa geometry to
a $+1$\% expansion in the final configuration, cf. Fig. \ref{structures}b
and c.

The only structural difference at the endpoint of the sliding motion
compared to the initially tested fcc/tetra-I geometry is that the O-Ru-O
trilayer as a whole is rotated by 60${}^o$ with respect to the metal
below, cf. Fig. \ref{structures}c and \ref{sliding}. While this leaves 
both the internal trilayer geometry, as well as the direct coordination
of the sub-surface oxygens to its Ru neighbours identical, the first
layer Ru atoms now end up in a stacking fault position. We find both
fcc/tetra-I geometries to be energetically and electronically degenerate
within our calculational uncertainty, so that the first layer Ru atoms (which
position with respect to the substrate is the only thing that changes)
cannot contribute noticeably to the bonding to the underlying metal
anymore. The only remaining coupling film-substrate occurs thus via
O${}_{\rm sub}$-Ru${}_{\rm 2nd}$ bonds, which are strongest in the
fcc/tetra-I geometry.

This sliding instability of the hcp/octa configuration also provides
a possible mechanism with which the most stable fcc/tetra-I geometry
can be realized: At first glance, it appears that kinetics will hinder 
the formation of the latter, as it seems to imply that all oxygen on-surface
adatoms in the O$(1 \times 1)$ phase, which initially occupy the hcp sites,
would have to change their position to fcc upon sub-surface O incorporation.
Instead of a collective jump, which would of course be penalized with a
high barrier, the penetration process could in reality and at finite
temperatures commence with finite-size islands of sub-surface oxygen
in all of the available sites (which are energetically not very different).
Above a certain critical extension, growing islands with the hcp/octa
configuration would then be liable to the sliding motion, once the 
significant energy gain of 0.2\,eV per O atom due to the shift overcomes
the cost due to edge effects (which are not considered in our periodic
calculations). This mechanism would be particularly facilitated at step
edges, where the lateral trilayer shift does not come into conflict
with the atoms of the neighbouring domain.

\begin{figure}
\epsfxsize=0.40\textwidth \centerline{\epsfbox{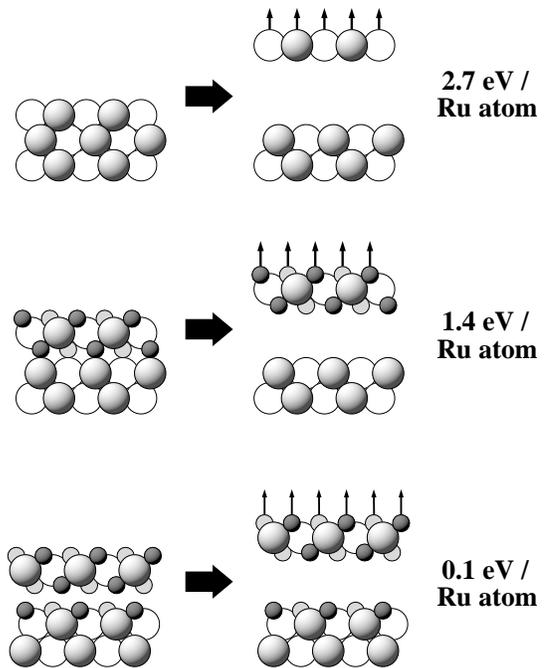}}
\caption{Energy per Ru atom required to lift off a complete
Ru layer from the Ru(0001) substrate. The strong decrease
along the sequence clean surface (top panel), 2 ML fcc/tetrahedral-I
(middle panel) and 3 ML floating trilayer (bottom panel)
indicates a significant destabilization of the Ru(0001) surface
upon increasing oxidation.}
\label{liftoff}
\end{figure}

Wrapping up the present argument, we believe that the key feature of
our calculations concerning the initial O incorporation into the metal
is the formation of two-dimensional $(1 \times 1)$ sub-surface O islands,
in which a Ru layer is getting nearly detached from the bulk through the
formation of an O${}_{\rm ad}$-Ru-O${}_{\rm sub}$ trilayer, which is more
strongly bound in itself rather than to the underlying substrate.
Although the sliding motion into the fcc/tetra-I geometry helps to enhance
this residual coupling of the trilayer to the substrate, it is
nevertheless significantly lower than that of the top Ru layer to
the substrate in the original pure metal. This is exemplified in
Fig. \ref{liftoff}, where we compare the energies per $(1 \times 1)$ 
cell required to completely lift off a metal layer in both cases: 
The original cost of 2.7\,eV per Ru atom to lift off a Ru layer from
clean Ru(0001) is almost halved, if the Ru layer is contained inside the 
O-Ru-O trilayer in the fcc/tetra-I structure (1.4\,eV / Ru atom), thus
reflecting a considerable destabilization of the metal surface upon
oxidation.

\subsection{Decoupling and continued oxidation}

From the understanding of a trilayer formation on top of Ru(0001), we
next turn to the question of how this film would continue to grow at
higher O coverages. As the lattice deformation cost renders sites
closest to the surface most stable and will again favour the formation
of dense O$(1 \times 1)$ islands, we address the continued oxidation
by directly calculating structures including a total of 3\,ML O, i.e. in
addition to the hitherto considered $(1 \times 1)$ on- and sub-surface
oxygen, we place another ML of O in sites between the second and third
substrate layer. Testing all combinations of the two possible on-surface
and three possible sub-surface sites totals to 18 trial geometries
at this coverage. Strikingly, the most stable geometry corresponds
to none of these, but is found, when the additional O ML is also
located between the first and second substrate layer, so that
both the tetra-I and the tetra-II sites are occupied as shown in
Fig. \ref{structures}d. 

Due to the additional sub-surface O, the O-Ru-O trilayer has thus not
grown in thickness itself, but has been almost completely decoupled
from the underlying Ru metal, as expressed by the tremendous first 
layer expansion of 116\% and the now almost zero binding per Ru
atom as shown in Fig. \ref{liftoff} \cite{note}. However, one should 
keep in mind that this virtual decoupling does not mean that the 
whole trilayer will lift off at finite temperatures: The binding of
0.1\,eV/Ru atom scales of course with the large number of Ru atoms
in a finite-size island, which would total to a high cost for
a complete detachment of the island. Yet, when comparing the binding
energies per Ru atom shown in Fig. \ref{liftoff}, the pronounced
destabilization trend along the oxidation sequence is very clear
and should be compared to the significantly lowered emission
temperatures of RuO${}_{\rm x}$ fragments from O-rich Ru(0001)
surfaces observed in recent temperature desorption spectroscopy 
(TDS) experiments \cite{boettcher00}.

With the O-Ru-O trilayer almost decoupled, the underlying substrate
will then primarily be influenced by the remaining third oxygen
ML sitting in tetra-II sites, i.e. in hcp sites with respect to
the second Ru layer. Indeed, when looking at the Ru substrate below
the trilayer in more detail, its electronic and geometric structure
very much ressembles a O$(1 \times 1)$ covered Ru(0001) surface
with 1\,ML O in hcp sites, cf. Fig. \ref{structures}a and d: The
trilayer on top has apparently only a negligible influence. Comparing
with the fcc/tetra-I configuration at 2\,ML coverage, the effect
of the additionally incorporated oxygen ML is thus to enable an
efficient decoupling of the O-Ru-O film by saturating the underlying
substrate bonds. 

It it interesting to compare these findings with
recently published data on the structure of ultrathin aluminium
oxide films on several metallic substrates \cite{jennison99}.
In that case a O$(1 \times 1)$ layer was also always found to reside 
below the oxide film and above the underlying metal, at a similar
position as in the corresponding chemisorption phase. Hence, the
presence of a terminal oxygen layer at the metal-oxide interface
(saturating the underlying substrate and in turn minimizing the
coupling to the oxide) could in fact be a more general phenomenon.

Given the negligible effect of the floating trilayer on the substrate
below, we finally deduce that the continued interaction of oxygen with
the Ru(0001) surface will proceed in an analogous fashion as before: In
the coverage sequence 3\,ML $< \theta <$ 4\,ML oxygen will be incorporated
below the Ru${}_{\rm 2nd}$ atoms leading to the formation of a second
O-Ru-O trilayer. The latter will then become decoupled by the fifth
ML oxygen, after which the next trilayer will be formed, and so on.
This way, a stack of weakly coupled trilayers would successively be
formed, below which a terminal sub-surface O layer saturates the metal
substrate.

\subsection{Accordion effect and transition to the oxide structure}

At this point it is natural to ask, whether the formation of the
O-Ru-O trilayers wouldn't already correspond to what is often termed
a surface oxide. Indeed, the sixfold coordination of Ru in the
trilayer together with the correct stoichiometry (one Ru, two O)
offers quite some resemblance to the bulk oxide, though one has 
to concede that the geometry is still distinctly different to the
rutile structure of the stable bulk RuO${}_2$.

\begin{figure}
\epsfxsize=0.48\textwidth \centerline{\epsfbox{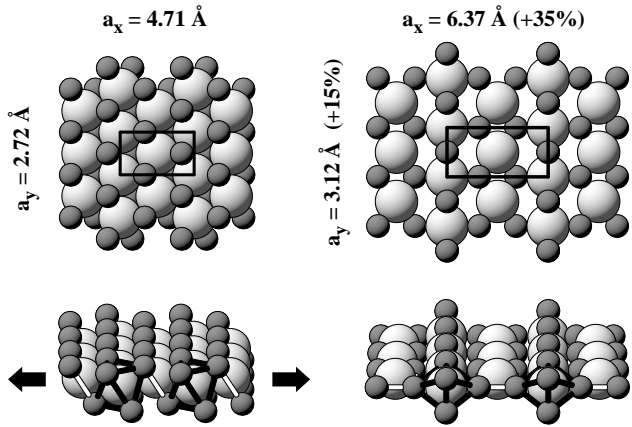}}
\caption{Atomic geometries of the O-Ru-O trilayer (left) and 
a RuO${}_2$(110) rutile layer (right). The top panels show a
top view, comparing the surface unit cells. The bottom panels
depict a perspective view, where O coordination octahedra
are drawn around the metal atoms. The trilayer can be
transformed into the rutile geometry by a lateral expansion
(along the arrows), keeping the drawn O-O bonds (white and black
lines) rigid. This way an alternating sequence of fully coordinated
Ru atoms (inside the tilted octahedra, black bonds) and
coordinatively unsaturated (cus) Ru atoms (surrounded by only
four O atoms, white bonds) is created.}
\label{accordion}
\end{figure}

Experimentally, the end product of the oxidation of the Ru(0001)
surface has been convincingly characterized as crystalline,
well-oriented RuO${}_2$(110) oxide domains, which are incommensurate
to the underlying Ru matrix, but aligned along the three $[{\bar{1}\bar{1}2}]$
directions on the (0001) basal plane \cite{over00}. Hence, we compare
in Fig. \ref{accordion} the structure of our O-Ru-O trilayer with
such a (110)-oriented rutile plane. At first glance, both have very
little in common and one notices primarily the largely different
dimensions of the surface unit cells, which obviously inhibit a
commensurate growth of RuO${}_2$(110) on a Ru(0001) substrate.
Yet, upon closer inspection it becomes clear, that the trilayer
can be transformed rather naturally into RuO${}_2$(110) by a
simple accordion-like lateral expansion, involving a tilting
of the fixed O octahedra around every second row of Ru atoms as
explained in Fig. \ref{accordion}. During this expansion the
self-contained trilayer, which offers the preferred sixfold O
coordination to every Ru atom, unfolds into a more open geometry,
in which every second Ru atom is now only fourfold coordinated,
i.e. so-called coordinatively unsaturated (cus) Ru atoms are formed.

\begin{figure}
\epsfxsize=0.48\textwidth \centerline{\epsfbox{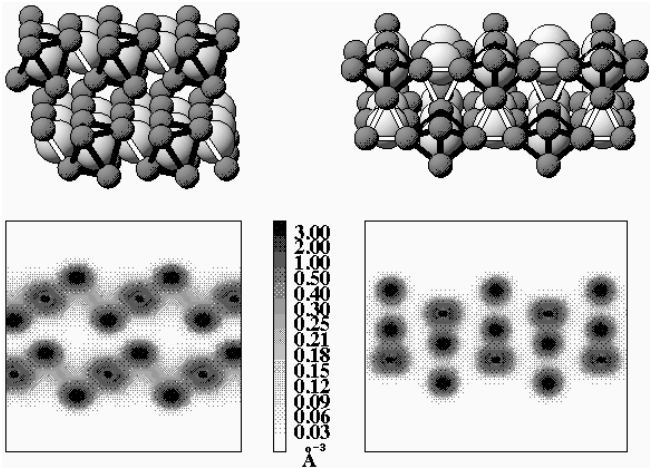}}
\caption{Comparison of the inter-trilayer bonding of the
O-Ru-O trilayer (left) and of RuO${}_2$(110) rutile layers 
(right): perspective view of the atomic geometry (top panels)
and calculated electron density perpendicular through the
slabs along the $[{\bar{1}\bar{1}2}]$ direction (bottom panels).}
\label{interlayerbonds}
\end{figure}

Although this accordion-effect offers thus a nice transformation
mechanism, without any material transport apart from mere
expansion, it is not immediately clear, what would drive the
self-contained trilayer into such a much more open and
coordinatively less favourable geometry. And in fact, the key
behind the transition becomes only apparent, when considering
higher film thicknesses: Whereas before the expansion two
trilayers are found to bind only negligibly to each other (which
is comprehensible, given that each trilayer alone already
offers a bulk oxide-like coordination to the metal atoms),
a stacking of rutile layers increases the local coordination
of the cus Ru atoms by the bridging O atoms of the next
layer, thus retrieving the ideal sixfold coordination for
all metal atoms as shown in Fig. \ref{interlayerbonds}. 

\begin{figure}
\epsfxsize=0.48\textwidth \centerline{\epsfbox{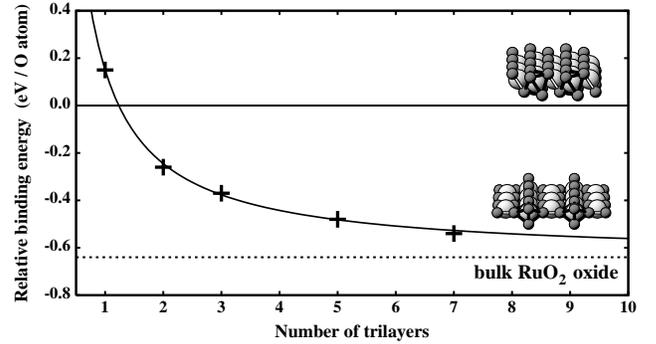}}
\caption{Relative binding energy difference between a stack
of O-Ru-O trilayers and of RuO${}_2$(110) rutile layers as
a function of layers in the stack. The negligible intertrilayer
bonding in the former case leads to a constant binding energy
per O atom, which is taken as the zero reference. Due to
the formation of new bonds the average binding energy per
O atom increases in the rutile case, rendering this structure
energetically more favourable than the O-Ru-O trilayers as
soon as two trilayers are packed onto each other. Additionally
drawn is the average binding energy in an infinite bulk
RuO${}_2$(110) arrangement (dotted line), which is the limit
approached by the rutile curve for thicker slabs.}
\label{trivsrutile}
\end{figure}

Precisely this additional inter-layer coordination in the
rutile case eventually leads to a stabilization of this structure
for higher film thicknesses. To this end, we compare in
Fig. \ref{trivsrutile} the relative binding energies of a
stack of trilayers with a corresponding stack of rutile layers,
where we neglect for the time being any residual influence
of the underlying Ru substrate. As the O-Ru-O trilayers
hardly bind to each other, the binding energy turns out
independent of the number of trilayers stacked onto each other.
On the contrary, the more rutile layers one packs together,
the more the average binding energy increases, indicating that
the increased coordination of the cus Ru atoms indeed leads
to the formation of inter-layer bonds. While one isolated
O-Ru-O trilayer is still by 0.15\,eV per O atom more stable
than its more open rutile counterpart, this situation reverses
already as soon as two trilayers are packed onto each other.
For thick rutile films, the binding energy finally approaches
gradually the limit, by which bulk rutile RuO${}_2$ is
more stable than an infinite stack of O-Ru-O trilayers, namely
by 0.64\,eV per O atom, cf. Fig. \ref{trivsrutile}.

Based on this comparison, we would hence estimate a phase
transition into a RuO${}_2$(110) oxide film after two
or more O-Ru-O trilayers have formed during the oxidation.
Yet, we notice that in this energetic comparison, the
residual influence of the underlying metal has been neglected.
While we have shown that this interaction is negligible
in the O-Ru-O trilayer case, cf. Fig. \ref{liftoff}, this
is not so clear for the incommensurable rutile layers on
Ru(0001). However, the exact value of this interaction,
which is limited to the atoms directly at the interface,
is not really relevant, as it will certainly not prevent
the transition to the rutile structure for thicker films, 
where the binding energy gain scales with the number of
inter-layer bonds formed throughout the whole film. Independent
of the interaction with the underlying metal substrate,
the film will therefore switch to the rutile structure
once a critical film thickness is exceeded, which concludes
our suggested oxidation pathway. The involved accordion-like
unfolding of the trilayers leads then rather naturally to 
RuO${}_2$(110) oriented domains, which are due to the large
lateral expansion incommensurate, but aligned to Ru(0001),
in agreement with the experimental data \cite{over00}.

In this respect it is interesting to notice that the calculated
energetic difference between such ultrathin RuO${}_2$ films and
the intermediate precursor configurations along the oxidation
pathway is not large. At elevated pressures and temperatures a
lively dynamics between these metastable configurations is 
therefore conceivable, which would of course affect the chemical
and mechanical properties of the surface. In the case of Ru, it
was already shown that contrary to common believe the oxidic
domains, which evolve on the surface in the reactive environment, 
form an essential ingredient to understand the catalytic activity
\cite{over00,boettcher97}. Yet, our findings suggest that the
situation might be even more complex, requiring possibly also
an inclusion of the dynamics between various metastable
configurations to achieve a reliable molecular-level description
of the catalytic reaction. To verify this for the particular
case of Ru and other metals (where we expect an analogous
situation), experimental studies addressing the microscopic
structure of surfaces in realistic environments (and/or under
steady-state reaction conditions) are obviously mandatory.

\subsection{Searching for the precursor}

As apparent from the preceding Sections, the salient feature
of our suggested oxidation pathway is the formation of an
O-Ru-O trilayer on top of Ru(0001) as a metastable precursor
to the final oxide film. Up to now, there is no direct
experimental evidence for this trilayer, which we attribute
to the fact, that all studies addressing the oxidation of the
Ru(0001) surface have hitherto unanimously employed rather
elevated temperatures, which in turn enable a rapid phase
transition to the final RuO${}_2$ bulk oxide structure. While 
it is presently not clear, whether it will at all be feasable 
to stabilize the trilayer by gently oxidizing the Ru(0001)
surface at sufficiently low temperatures, we still hope
that our results may inspire carefully directed experiments 
by listing in the following a number of observables that might
be used as a fingerprint.

Along our suggested oxidation pathway, the O-Ru-O trilayer 
occurs in three different positions with respect to the underlying
metal substrate: first in a hcp/octa configuration, then
in a fcc/tetra-I geometry after the surface registry shift and
finally decoupled by the third incorporated O layer. From a 
structural point of view, all three precursor geometries
are laterally still commensurate to the underlying Ru(0001).
It is only the significant expansion during the accordion-unfolding
that finally leads to the distinctly different RuO${}_2$(110)
lattice  spacing of the oxide domains. Just up to this last
transition, the surface should therefore only exhibit the hexagonal 
$(1 \times 1)$ low energy electron diffraction (LEED) pattern
corresponding to the basal Ru(0001) plane. Whether or not
the trilayer has switched to the final rutile structure can
therefore easily be monitored by the appearance of new LEED 
spots, which due to the incommensurability of both lattices
show up at completely different positions on the screen
\cite{kim00}. Samples, for which TDS curves tell that more 
than one oxygen ML have been deposited and for which a sharp
$(1 \times 1)$ LEED pattern without extra spots is observed,
are therefore most likely candidates for sub-surface oxygen.

If such a preparation has led to the formation of $(1 \times 1)$
sub-surface islands, could then be visible in scanning tunneling
microscopy (STM) experiments: The sub-surface oxygen induced 
vertical relaxations are huge, cf. Fig. \ref{structures}, and
should enable an identification of the corresponding islands in
the STM images. With respect to the surface registry shift,
it would be particularly interesting to check whether the atomic
lines in the $(1 \times 1)$ periodic islands are still in
registry with the surrounding domains across the island perimeter: 
After the surface registry shift, the on-surface oxygens in the
island will be in fcc positions in contrast to the hcp site
in the neighbouring O$(1 \times 1)$ domains, which should
correspondingly show up in an STM image. Of course, the
largely different layer distances in the sub-surface patches
will also lead to distinctly different LEED I(V) curves,
which could therefore also be used for fingerprinting. Even
with only poorly defined long range order on the surface, the 
significantly increased Ru-Ru bondlengths between the metal
atoms in the trilayer and the metal atoms of the underlying
substrate might still show up in extended x-ray absorption
fine structure (EXAFS) data.

\begin{figure}
\epsfxsize=0.48\textwidth \centerline{\epsfbox{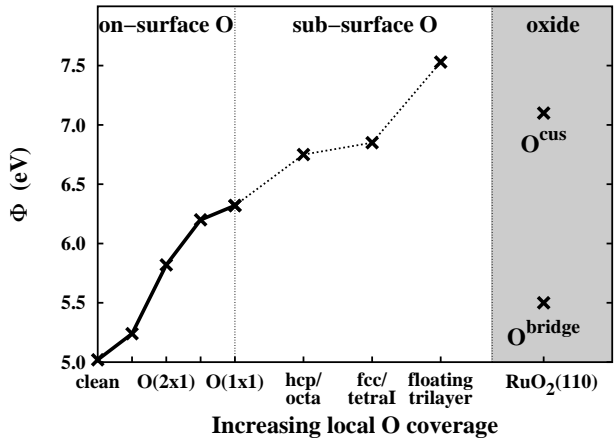}}
\caption{Calculated workfunction, $\Phi$, along the oxidation pathway.
The work-function rises throughout the on-surface chemisorption
of O at $0 < \theta \le 1$\, ML, but is even increased at the
three metastable precursor configurations, cf. Fig. \ref{structures}.
For comparison also the calculated workfunction of the two possible 
terminations of RuO${}_2$(110) domains is shown. The latter oxidation 
stage (in the grey shaded area) should be clearly distinguishable
from the sub-surface O precursors by the appearance of new LEED
spots as described in the text.}
\label{workfunction}
\end{figure}

In Fig. \ref{workfunction} we show the computed workfunctions,
$\Phi$, for the geometries along the oxidation sequence. Up to
the completion of the full $(1 \times 1)$ on-surface adlayer,
$\Phi$ rises continuously, which has been explained in terms of
the increasing dipole moment due to the chemisorbed oxygen
\cite{stampfl96b}. Interestingly and contrary to common believe,
we find that the incorporation of O into the sub-surface
region does not cause an analogous reduction of the workfunction
due to the build-up of a now inverted dipole moment. Instead,
$\Phi$ continues to rise inside the $(1 \times 1)$ islands, until
it reaches a very high value of 7.53\,eV for the floating trilayer
structure. In Fig. \ref{workfunction} we also included the
workfunctions for the two possible surface terminations of
the final RuO${}_2$(110) oxide domains \cite{kim01,reuter01b}: In
addition to the traditionally conceived stoichiometric rutile(110)
termination, O${}^{\rm bridge}$, oxygen-rich conditions can further
stabilize a so-called polar termination, O${}^{\rm cus}$, with
excess terminal O atoms, which lead to a significantly higher
work function compared to the stoichiometic case \cite{reuter01b}.
In light of the results shown in Fig. \ref{workfunction}, we
argue that the decrease in the workfunction upon oxidation
reported by B\"ottcher and Niehus \cite{boettcher99} (which was then
attributed to the presence of sub-surface oxygen) in fact reflected
already the gradual formation of O${}^{\rm bridge}$ oxide domains 
on the surface, whereas sub-surface oxygen would in reality lead
to increasing workfunctions. Note, that the oxide formation goes
hand in hand with the aforementioned appearance of new LEED spots,
so that a high workfunction due to the formation of O${}^{\rm cus}$ 
terminated domains under oxygen-rich preparation conditions should
still be clearly distinguishable from a high workfunction due to the
floating trilayer. Concomitantly, we suggest that an increasing
work function for coverages $\Theta > 1$\,ML without the appearance
of new LEED spots would represent a likely (yet rather ambiguous)
fingerprint for the O-Ru-O trilayer. To better take the heterogeneity
of a surface, which includes sub-surface O islands, into account,
we therefore believe photoemission of adsorbed xenon (PAX) experiments
measuring the local workfunction to be a more suitable
technique. 

Although X-ray Photoemission Spectroscopy (XPS) is in principle
also a most eligible tool to study the oxidation of metal surfaces, 
we unfortunately find in the present case that it would not lead to 
a clearcut signal, allowing to identify the precursor geometries.
In a recent publication, we have described that in the sequence of
ordered adlayers the Ru $3d$ level shifts by up to 1\,eV towards
higher binding energies, while the O $1s$ level remains almost
unchanged \cite{lizzit01}. In contrast, the formation of the oxide 
domains is then characterized by only a small back shift of the Ru
$3d$ core level towards lower binding energies compared to its position
in the coexisting O$(1 \times 1)$/Ru(0001) phase, while it is now the O
$1s$ which largely shifts in the same direction \cite{reuter01c}. To assess 
the XPS signal received from the possible intermediate precursors,
we calculated all Ru $3d$ and O $1s$ surface core level shifts (SCLS)
for the hcp/octa, fcc/tetra-I and floating trilayer geometries
inside the sub-surface O islands, cf. Fig. \ref{structures},
using exactly the methodology described in detail in refs.
\onlinecite{lizzit01,reuter01c}. Unfortunately, the Ru $3d$ levels
in all three structures turn out to lie very close to the position 
in the coexisting O$(1 \times 1)$/Ru(0001) phase, so that the 
incorporation of sub-surface O would not lead to an unambiguously
identifyable new signal, but at best to a small shoulder aside an
existing peak. This situation is different for the O $1s$ levels,
which particularly for the floating trilayer are shifted by 1.1\,eV
towards lower binding energy compared to the position in the
chemisorption phases. However, there they almost coincide with the
O $1s$ levels of the RuO${}_2$(110) domains \cite{reuter01c}, so
that the presence of such a peak would only be a clear fingerprint
for the sub-surface precursor, if one can safely rule out that no 
oxide domains have been formed yet.

\begin{figure}
\epsfxsize=0.43\textwidth \centerline{\epsfbox{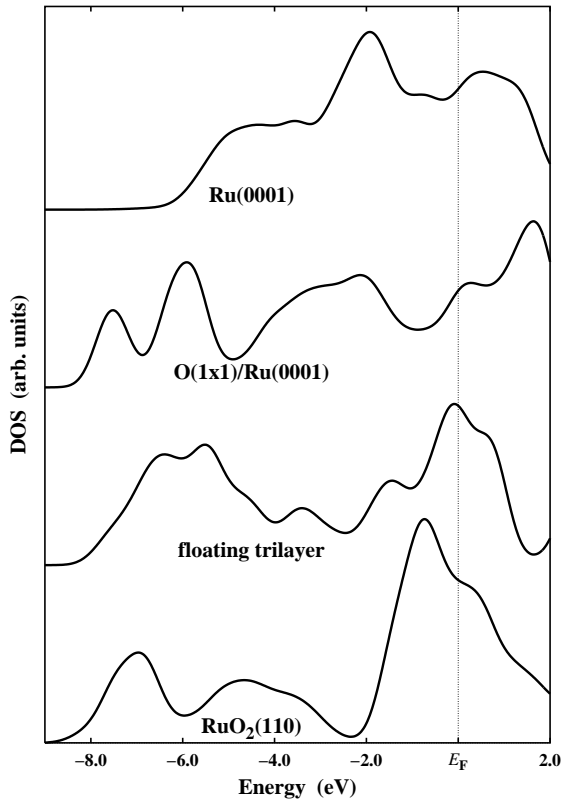}}
\caption{Calculated $4d$ DOS of first layer Ru atoms in
various geometries along the oxidation pathway. The DOS of
the first layer Ru atoms in the hcp/octa and fcc/tetra-I positions
of the O-Ru-O trilayer (not shown) are very similar to the
one shown here, where the trilayer is already almost decoupled
from the underlying Ru(0001) substrate at a coverage of
$\theta = 3$\,ML. The curves have been shifted and strongly
smoothed for clarity.}
\label{dos}
\end{figure}

This ambiguous situation between precursor and final oxide
structure is similar in the case of ultraviolet photoemission
spectroscopy (UPS). There the disappearance of a broad
feature at about -2.5\,eV with respect to the Fermi-energy,
$E_{\rm F}$, was suggested as a marker for the oxidation of
the Ru(0001) surface \cite{boettcher99b}. Well aware of the fact,
that calculated densities of states (DOS) can only cautiously
be used to interpret UPS data, we show in Fig. \ref{dos} how 
this quantity evolves for the first layer Ru atoms (which
would contribute most to the UPS signal) along the oxidation 
pathway. The DOS for the clean surface shows the Ru $4d$ band,
which is slightly narrower than the bulk band due to the lowered
coordination of the surface atoms \cite{lizzit01} and which
is dominated by a peak at -2.0\,eV. Analyzing the electron
density distribution of the states that contribute to this peak,
we deduce that the latter is due to backbonding states to the
second substrate layer. The formation of chemisorption states
at the edges of the $4d$ band can clearly be seen in the DOS of
the O$(1 \times 1)$ on-surface phase and goes hand in hand
with a reduction of the feature around -2.0\,eV, which according
to the electron density distribution analysis is still due to
backbonding states to the second layer. The depletion of such
states agrees well with the 5\% expansion of the first layer
distance upon O adsorption as shown in Fig. \ref{structures}a.
We notice in passing, that the bonding oxygen-induced states at
the bottom of the band are split into two peaks due to the
different lateral interaction of states of $p_x/p_y$ and $p_z$
symmetry within the densely packed O overlayer. Addressing
next the DOS of the floating trilayer precursor shown in Fig.
\ref{structures}d, we find now almost no states left at
the energies around -4.0 to -2.0\,eV reflecting the fact, that
the floating trilayer has virtually no Ru-Ru backbonding
to the underlying substrate anymore. Continuing to the DOS
of the RuO${}_2$(110) domains, this lack of direct Ru-Ru bonds
in the oxide leads similarly to few states around this energy
region. Instead, the predominant features appear now at
-7.0 to -4.0\,eV due to O $2p$ - Ru bonding states and 
slightly below the Fermi level \cite{sorantin92}. Based
on these results, we suggest that the depletion of the UPS
feature around -2.5\,eV, which had been suggested as an
oxidation marker, simply indicates the diminishing Ru-Ru 
backbonds at the oxidizing surface, but would not allow to 
distinguish between the intermediate precursor and the
final oxide domains.

\section{Summary}

In conclusion, we have presented an atomistic pathway
leading from a clean Ru(0001) surface to the RuO${}_2$(110)
domains, that were experimentally reported as the end
product of the oxidation of this Ru surface. Oxygen penetration
into the lattice starts only after the full chemisorbed
adlayer is completed on the surface. The ensuing oxygen
incorporation leads to the formation of two-dimensional
sub-surface O islands between the first and second metal
layer, in which an O-Ru-O trilayer gets decoupled from
the underlying substrate. Continued oxidation successively
leads to the formation of a stack of such trilayers, which 
finally unfold into the RuO${}_2$(110) oxide structure once
a critical film thickness is exceeded.

The salient feature of this suggested oxidation pathway
is a floating O-Ru-O trilayer as a metastable precursor
to the final oxide film. Detailed structural and electronic
data characteristic of this precursor was given in order
to inspire and enable a directed experimental search for
it at low oxidation temperatures. The large lattice relaxations
induced by the sub-surface oxygen render structural techniques
like STM, LEED and EXAFS as most promising techniques for such 
a search, while the precursor would not lead to clearcut 
fingerprints in XPS and UPS data. A rising workfunction
beyond the completion of the adlayer together with a 
$(1 \times 1)$ LEED pattern would be a positive signal for 
the predicted intermediate, where particularly PAX measurements
of the local workfunction would be ideally suited to take the
heterogeneity of the oxidized surface into account. We find 
all intermediate precursors along the oxidation pathway to be
rather close in energy to the final oxide structure, which 
indicates a likely lively dynamics between the metastable
configurations under realistic conditions and which might affect
the chemical and mechanical properties of the surface.

\section*{Acknowledgements}

We gratefully acknowledge stimulating discussions concerning
the experimental studies on oxidized Ru(0001) with A. B\"ottcher,
B. Krenzer, R. Blume and H. Conrad. This work was partially
supported by the Deutsche Forschungsgemeinschaft (Schwerpunkt
``Katalyse'').


\begin{references}
\bibitem{note}
Note, that due to the negligible coupling of the trilayer to the
underlying substrate, it is irrelevant whether we put it in a
stacking fault geometry or not: both cases are electronically,
structurally and energetically absolutely degenerate.
\end{references}
\end{document}